\documentclass[review]{elsarticle}

\usepackage{subfig}
\usepackage{multirow}
\usepackage{enumitem}
\usepackage{color}
\usepackage{amsmath}
\usepackage{lineno}
\usepackage{array}
\usepackage{placeins}
\usepackage{changepage}
\definecolor{corr}{rgb}{0,0,0}
\newcommand{\modif}[1]{\textcolor{corr}{#1}}

\journal{Artificial Intelligence in Medicine}

\begin{document}
\begin{frontmatter}

\title{Offline identification of surgical deviations in laparoscopic rectopexy}

\author[label1,label2]{Arnaud Huaulm\'{e}}
\author[label2]{Pierre Jannin\corref{cor1}}
\ead{pierre.jannin@univ-rennes1.fr}
\author[label4]{Fabian Reche}
\author[label4]{Jean-Luc Faucheron}
\author[label1,label5]{Alexandre Moreau-Gaudry}
\author[label1]{Sandrine Voros}

\cortext[cor1]{Corresponding author}

\address[label1]{UGA / CNRS / INSERM, TIMC-IMAG UMR 5525, Grenoble, F-38041, France}
\address[label2]{Univ Rennes, INSERM, LTSI - UMR 1099, F35000, Rennes, France}
\address[label4]{Colorectal Unit, Department of Surgery, Michallon University Hospital, F-38043 Grenoble, France}
\address[label5]{UGA / CHU Grenoble / INSERM, Centre d'Investigation Clinique - Innovation Technologique, CIT803, Grenoble, F-38041, France}

\begin{abstract}
{\bf Objective:} According to a meta-analysis of 7 studies, the median number of patients with at least one adverse event during the surgery is 14.4\%, and a third of those adverse events were preventable. The occurrence of adverse events forces surgeons to implement corrective strategies and, thus, deviate from the standard surgical process. Therefore, it is clear that the automatic identification of adverse events is a major challenge for patient safety. In this paper, we have proposed a method enabling us to identify such deviations. We have focused on identifying surgeons' deviations from standard surgical processes due to surgical events rather than anatomic specificities. This is particularly challenging, given the high variability in typical surgical procedure workflows.

\noindent{\bf Methods:} We have introduced a new approach designed to automatically detect and distinguish surgical process deviations based on multi-dimensional non-linear temporal scaling with a hidden semi-Markov model using manual annotation of surgical processes. The approach was then evaluated using cross-validation.

\noindent{\bf Results:} \modif{The best results have over 90 \% accuracy. Recall and precision for event deviations, i.e. related to adverse events, are respectively below 80 \% and 40 \%. To understand these results, we have provided a detailed analysis of the incorrectly-detected observations}

\noindent{\bf Conclusion:} Multi-dimensional non-linear temporal scaling with a hidden semi-Markov model provides promising results for detecting deviations. Our error analysis of the incorrectly-detected observations offers different leads in order to further improve our method.

\noindent{\bf Significance:} Our method demonstrated the feasibility of automatically detecting surgical deviations that could be implemented for both skill analysis and developing situation awareness-based computer-assisted surgical systems.
\end{abstract}

\end{frontmatter}

\section{Introduction}
\label{intro}

In the review \cite{anderson_surgical_2013}, the authors have identified 7 publications between 1991 and 2008 reporting adverse events (AEs). Over all these studies, the median number of patients having undergone one or several AEs was 14.4\%, and a third (37.9\%) of those AEs were considered preventable. An AE is defined by the World Health Organization (WHO) as ``an injury related to medical management, in contrast to complications of disease. Medical management includes all aspects of care, including diagnosis and treatment, failure to diagnose or treat, and the systems and equipment used to deliver care'' \cite{world_health_organization_who_2005}. In the surgical field, we can distinguish between different events categorized as postoperative adverse events (pAEs) for AEs occurring following surgery, and intraoperative adverse events (iAEs), when AEs occur during surgery.

Hospitals use risk management to prevent AEs. This consist in identifying and characterizing AEs along with their severity, with the aim to propose strategies designed to reduce the likelihood they will occur again. The identification consists in determining when an AE occurred and which anatomic structure was affected. The characterization consists in determining the AE's severity. However, since both steps are performed manually, this is a costly and time-consuming process, prone to both subjectivity and mistakes. In this paper, we have analyzed the relevance of surgical process models (SPMs) to help identification of iAEs.

A SPM is ``a simplified pattern of a surgical process that reflects a predefined subset of interest of the surgical process in a formal or semi-formal representation'' \cite{jannin_modeling_2001}. A SPM describes a surgical procedure at different granularity levels: phases, steps, and activities \cite{lalys_surgical_2013}. A surgical procedure is divided into successive phases corresponding to the procedure's main periods. A phase is composed of one or several steps. A step is a sequence of activities deployed to achieve a surgical objective. An activity is a physical action performed by the surgeon. Each activity is deconstructed into different components, including the action verb, anatomic structure concerned by the action, and surgical instrument employed to perform this action.

Surgical process modeling has been used in various applications, such as surgical skills evaluation \cite{riffaud_recording_2010,forestier_classification_2012}, operating room management optimization \cite{sandberg_deliberate_2005, padoy_-line_2008}, or robotic assistance \cite{ko_surgery_2007,nomm_recognition_2008}. However, SPMs have rarely been applied for surgical quality assessment. A method was presented in \cite{bouarfa_workflow_2012} for detecting modifications from the standard process called deviations. The authors employed surgical tool information  to create a standard surgical process and draw correlations between this standard surgical process and a specific surgery using the Needleman-Wunsch global alignment algorithm. One limitation of this method is that the reasons for the deviations are not identified. 

Deviation detection has been studied in other domains, such as bank \cite{jadhav_anomaly_2013} or software security \cite{tan_hidden_2008}. The principle of deviation detection relies on constructing a standard process and detecting deviations using a comparison between this standard process and a new one. To the best of our knowledge, these authors did not distinguish different types of deviations either.

To overcome this limit, we propose for the surgical field, as illustrated in table~\ref{tab:deviations}, three types of surgical deviations based on observations from participating surgeons. It is important to note that the notion of deviation is independent of the occurrence of AE, this notion only reflects the modification of the standard surgical process model. An AE could occur as a result of any type of deviation or even if no deviation is visible.

\begin{table*}[!ht]
	
	\centering
	\caption{Surgical deviation types and definitions. }
    \begin{tabular}{>{\centering\arraybackslash}m{.20\linewidth}|>{\centering\arraybackslash}m{.70\linewidth}}
         \shortstack{Surgical \\deviation type}   &  Definition  \\\hline
         Context deviations & Deviations due to patient's particularities as anatomic specificities, patient's pathology, and co-morbidity; this category also considers all deviations due to the surgical context, as operating room disruptions. \\\hline
         Expert deviations &  Deviations due to the surgeon who performs the surgery; this category includes deviations due to surgical expert knowledge, and surgeons’ habits or preferences.\\\hline
         Event deviations & Deviations from the usual surgical process to correct or limit the impact of iAEs.
    \end{tabular}
    \label{tab:deviations}
\end{table*}

To help the identification of iAES, this work aims to detect surgical deviations from a standard surgical process and classify them according to the above categories. For this purpose, we propose using an extension to non-linear temporal scaling \cite{forestier_non-linear_2014}, called Multi-Dimensional Non-Linear Temporal Scaling (MD-NLTS), with the aim to detect deviations and a hidden semi-Markov model (HsMM) designed to classify them.

\section{Material and methods}
\label{Method}
\begin{figure*}[h]
 \centering
 \includegraphics[width=\textwidth]{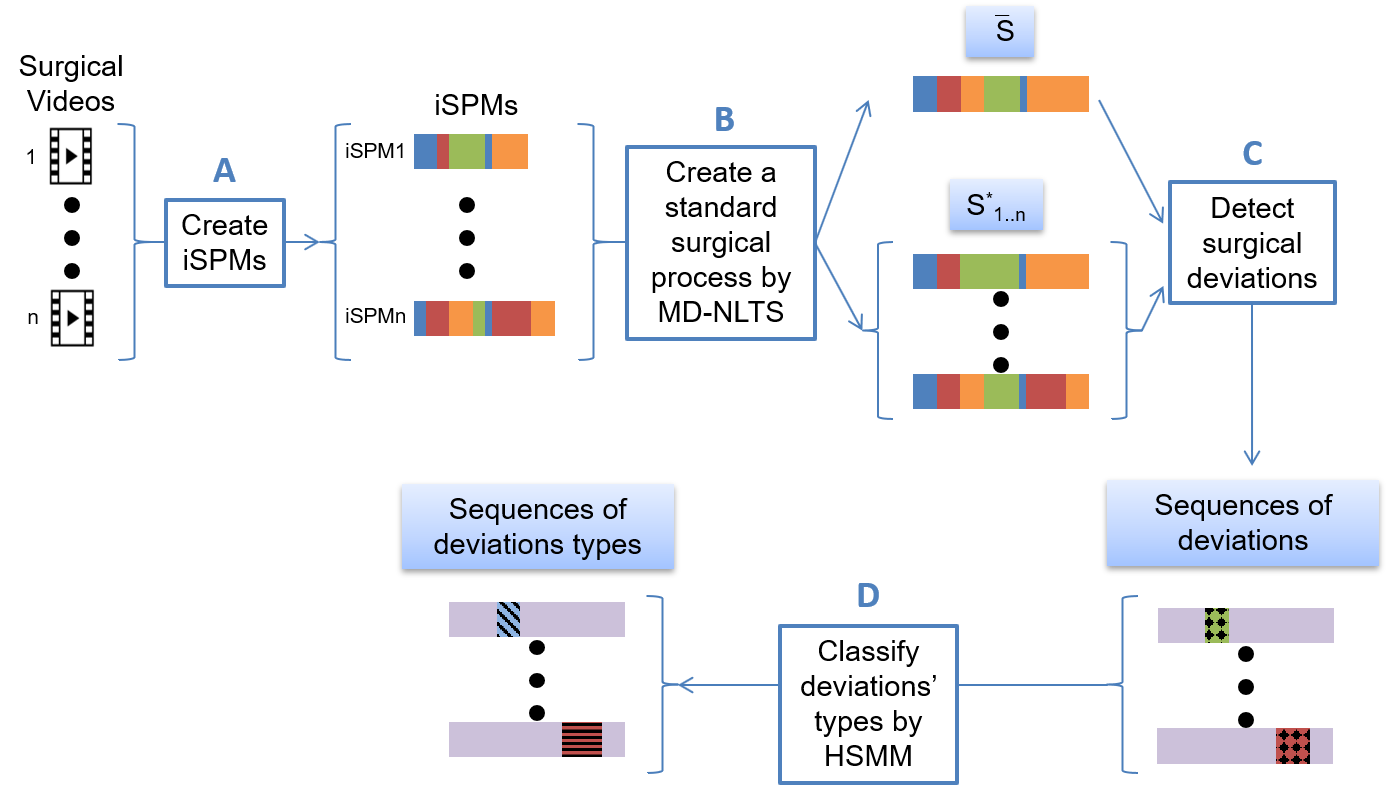}
 \caption{{\bf Classification of surgical deviation types based on four modules.}
Module A enables the creation of individual Surgical Process Models (iSPMs) based on surgical video annotations. Each color in the iSPMs represents one type of activity. Module B provides two outputs, a standard surgical process $\overline{S}$, and the sequences $S^*_{1..n}$ temporally aligned to $\overline{S}$. Module C compares one aligned sequence to the standard surgical process $\overline{S}$ for each instant to detect a potential deviation. The deviations are highlighted by dots in deviation sequences. Module D classifies each type of surgical deviation, i.e., blue diagonal crosshatch for a context deviation and red horizontal crosshatch for an event deviation.}
 \label{fig:method}
\end{figure*}

This section presents our offline method to detect and classify deviations in rectopexy surgery for skill analysis, as summarized in Fig~\ref{fig:method}. 

Our method is composed of four modules: A) the creation of individual surgical process models (iSPMs) based on clinical data; B) the creation of a standard surgical process; C) the detection of surgical deviations; D) the classification of deviation types. Each of these modules is described in the following subsections.

\subsection{Creation of individual surgical process models based on clinical data (Module A)}

The objective of this module is to describe surgical procedures with individual surgical process models (iSPMs) based on observations of surgical videos.

\subsubsection{Data}
\label{Data}
The dataset used in this paper consists of 11 endoscopic videos of laparoscopic rectopexies. A rectopexy is a digestive surgery that consists of correcting the anal prolapse by fixing the rectum to the sacrum through meshes. The operations were performed by a single expert surgeon at the Grenoble University Hospital, France, involving 11 women who had not undergone a hysterectomy during previous hospitalization.

This study was approved by an ethics committee and declared to the French authorities (CCTIRS\footnote{Comit\'e Consultatif sur le Traitement de l'Information en mati\`ere de Recherche dans le domaine de la Sant\'e.} and CNIL\footnote{Commission Nationale de l'Informatique et des Libert\'es.}). All patients operated by the participating surgeon between January 2015 and December 2017 were included as long as they met the declared inclusion criteria and signed a written informed consent authorizing data collection and data utilization for this study.

Since the surgeon performed the surgeries in a limited period, we considered that his knowledge, habits, and preferences did not vary enough to introduce expert deviations. We considered that the dataset thus contained only context and event deviations. However, some rare variations of performance of an expert may occur due to personal reasons. This was not considered here, but this could be checked in a prospective study.

\subsubsection{Creation of individual surgical process models}
\label{iSPM}
For the creation of iSPMs, we have focused on the two following phases: dissection and resection. The objectives of these phases were to respectively access both fixation points (rectum and sacrum) and remove the Pouch of Douglas. According to the participating surgeons, these phases are the most difficult to perform and most likely to cause AEs. In another hand, these phases are the most standardized according to patients’ particularities. One element able to modify the process is the presence or not of a uterus. In our dataset, no patient underwent a previous hysterectomy.

To create iSPMs, first, a Cognitive Task Analysis (CTA) \cite{clark_cognitive_2008} was conducted by a bio-medical engineer familiarized with this methodology and involved two expert surgeons including the one who performed the surgical procedures. The objective of the CTA was to capture and understand the expert knowledge to allow the annotation of the activities and iAEs. The latter were manually recorded by the bio-medical engineer thanks to the ``Surgery Workflow Toolbox [annotate]" software \cite{garraud_ontology-based_2014}. The annotation of the 11 surgeries represent a total of 671 activities (e.g. cut the rectum with a monopolar hook), and 16 iAEs. All of these iAEs were bleeding events, some on which lasting only a few seconds, though others over two minutes. Each of them was validated by an additional surgeon who did not perform the surgeries.

At this point, the iSPM is a label sequence composed of a succession of activities defined by three components (action verb, surgical instrument, and anatomic target) \cite{lalys_surgical_2013} and characterized by a duration. Thus, each iSPM is a continuous sequence of activities characterized by a duration (see Module A in Fig~\ref{fig:method}).

\subsection{Creation of a standard surgical process by multi-dimensional non-linear temporal scaling (Module B)}\label{subsec:creationRef}
The objective of this module is to create a standard surgical process that represents the most typical sequence of activities performed by the surgeon, to be used for detecting deviations in the third module. This second module is composed of three steps: 1) sampling the iSPMs; 2) aligning the sampled iSPMs to get the same length for all iSPMs; 3) creating the standard surgical process itself.

\subsubsection{Sampling the individual surgical process models}
As explained previously, iSPMs are continuous sequences. However, to perform the following steps, we need discrete sequences. We thus sampled the iSPMs to achieve this goal. The impact of the sampling rate is analyzed in the validation section.

\subsubsection{Alignment of iSPMs by multi-dimensional non-linear temporal scaling}
\label{MD_NLTS}
To create a standard surgical process according to activity sequences rather than their durations, we needed to temporally align the iSPMs. To this end, we have proposed a new approach called Multi-Dimensional Non-Linear Temporal Scaling (MD-NLTS), inspired by the Non-Linear Temporal Scaling (NLTS) proposed in \cite{forestier_non-linear_2014}.

NLTS is a multiple alignment method developed for one-dimensional surgical processes alignment. It is derived from dynamic time warping (DTW) and involves three steps: 
\begin{enumerate}[label=\alph*)]
	\item An average sequences of the set of sequences is computed by DTW Barycenter Averaging (DBA) \cite{petitjean_global_2011};
	\item The average sequence is independently aligned to each sequence of the set, with the aim of defining which elements of the sequence correspond to each element $l$ of the average sequence. Thus, for each element $l$ we have the corresponding set of elements of all sequences and $widths[l]$ and the maximum number of elements in the set sequence corresponding to each element $l$ of the average sequence;
	\item The alignments are finally ``unpacked'': All sequences are warped to include the same number of elements, defined by $widths[l]$, in a way that avoids information loss. 
\end{enumerate}

NTLS was created to overcome one DTW limitation. To perform multiple alignments with DTW, one sequence must be chosen as the reference, with the other sequences aligned to this reference. The alignment is thus dependent on the chosen reference. On the contrary, NLTS enables alignment between three or more sequences by computing an average sequence using DBA \cite{petitjean_global_2011}. NLTS realizes a local alignment by focusing on regions with string similarity rather than on all sequences' durations. Moreover, with this alignment, there is no loss of information, given that the sequences are extended during step c of NLTS, so that even an item with few samples will be retained anyway.

Despite these advantages, NLTS only enables the alignment of one-dimensional surgical processes. Thus, when we seek to align activity sequences in which activities are composed of three components (action verb, surgical instrument, and anatomic target), NLTS considers these three components as one dimension. If we have three activities, defined as follows: A\_1: \textless verb\_1, instrument\_1, target\_1\textgreater, A\_2: \textless verb\_1, instrument\_2, target\_1\textgreater{} and A\_3: \textless verb\_3, instrument\_3, target\_3\textgreater, NLTS will consider A\_2 and A\_3 equally different from A\_1, even though the instrument only differs between A\_1 and A\_2.

In \cite{shokoohi-yekta_non-trivial_2015}, two approaches were proposed to achieve multi-dimensional warping: either dependent warping or independent warping. However, these approaches were only applied to classic DTW. Thus, to take into account the benefits of multi-dimensional warping and NLTS, we propose a dependent multi-dimensional warping applicable to NLTS. We chose to develop a dependent warping approach, because the three components are strongly linked within the activity.

We adapted the NLTS cost matrix, used in step a) of NLTS, to develop MD-NLTS. In NLTS, the cost matrix between sequence $Q$ and sequence $C$ is defined as:

\begin{equation}
    d(q_{i},c_{j})=
    \begin{cases}
      0, & \text{if}\ q_{i}=c_{j} \\
      1, & \text{if}\ q_{i} \neq c_{j}
    \end{cases}
\end{equation}

where $q_{i}$ is the label of sequence $Q$ at $t=i$, and $c_{j}$ the label of sequence $C$ at $t=j$.

In MD-NLTS, each sequence is composed of $M$ dimensions. We define element $D$ of the cost matrix as the sum of the distance of each dimension:

\begin{equation}
    D(q_{i},c_{j})=\sum_{m=1}^{M} d(q_{i,m},c_{j,m})
    \label{equ:MD-NLTS}
\end{equation}
with,

\begin{equation}
    d(q_{i,m},c_{j,m})=
    \begin{cases}
      0, & \text{if}\ q_{i,m}=c_{j,m} \\
      1, & \text{if}\ q_{i,m} \neq c_{j,m}
    \end{cases}
\end{equation}

For the three activities, A\_1, A\_2, and A\_3,  previously defined, MD-NLTS will consider A\_1 more similar to A\_2 ($D(_{A\_1,A\_2})=1$) than A\_3 ($D(_{A\_1,A\_3})=3$). This difference will impact step a) by influencing the average sequence created by DBA. The other MD-NLTS steps are similar to NLTS. Following the alignment, all aligned sequences $S_{1..n}^*$ exhibit the same length (see Module B in Figure~\ref{fig:method}).

\subsubsection{Computation of the standard surgical process} \label{creationSSP}
The standard surgical process $\overline{S}$ is created by computing the more frequent activity in all aligned sequences $S_{1..n}^*$, at each instant. Let’s assume that we have the following activities at instant t: $s_1^* [t]= A\_1$, $s_2^* [t]= A\_1$ and $s_3^* [t]= A\_2$. Activity $A\_1$ is more frequent than $A\_2$, so $\overline{S}[t] = A\_1$.

Although MD-NLTS computes an average sequence in its first step, we did not select this as the standard surgical process given that this average sequence does not have the same length as the aligned sequences, rendering it impossible to detect deviations by comparing the activities. The coherence of this proposed standard surgical process was validated with the surgeons (see Section \ref{validationStdSurgPro}).

\subsection{Detection of surgical deviations (Module C)}
The objective of this third module is to detect deviations by comparing an aligned sequence $s$ of $S_{1..n}^*$ to the standard surgical process $\overline{S}$. To compare these sequences, we compute the distance $D(\overline{S}_{t},S^*_{s,t})$ between these two surgical processes at each time-step $t$. Contrary to the computation of $D$ in Eq~(\ref{equ:MD-NLTS}), the distance between the two sequences is computed for the same time-step since the surgical sequences are aligned. Similarly to step \ref{MD_NLTS}, the distance is multidimensional, i.e. the three components of the sequence are taken into account. Deviations are detected at each instant $t$ when $D(\overline{S}_{t},S^*_{s,t})>0$.

\subsection{Classification of deviation types by a hidden semi-Markov model (Module D)}
\label{HsMM}
The objective of a hidden semi-Markov model (HsMM), also called explicit-duration HMM \cite{yu_efficient_2003}, is to explain a non-observable sequence (i.e., the hidden state sequence) using an observable sequence (i.e., the observation sequence) in which it is theoretically conceivable to stay in the same state for an infinite duration. An HsMM is characterized by $\lambda = (\pi, A, B, P)$, where:
\begin{enumerate}
 \item $\pi$ is the initialization matrix containing probabilities to start the sequence at each state;
 \item $A$ is the transition matrix between hidden states containing the probabilities of changing states between two instants. Self-transitions are impossible ($A_{i,i}=0$, for each $i\in[0, nb\_hidden\_state]$);
 \item $B$ is the emission matrix containing the probabilities of producing given observations knowing that we are in a specific state;
 \item $P$ is the state duration matrix defining the probability to stay in a specific state for each possible duration.
\end{enumerate}

\begin{figure}[h]
  \subfloat[$D(\overline{S},S_s^*)$ : Distance between a specific surgery and the standard surgical process $\overline{S}$. Blue=0, yellow=1, grey=2, red=3.]{
    \includegraphics[width=\linewidth,height=0.2cm]{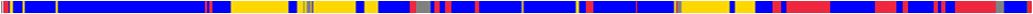}
    \label{distToMean}
  }\newline
  \subfloat[Event in a specific surgery. Blue= no event, yellow = intraoperative adverse event.]{
    \includegraphics[width=\linewidth,height=0.2cm]{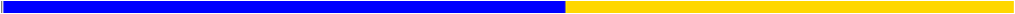}
   \label{event}
  }\newline
  \subfloat[True hidden state. Blue=no deviation, yellow= context deviation, grey= event-deviation.]{
    \includegraphics[width=\linewidth,height=0.2cm]{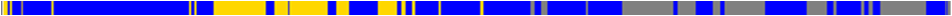}
    \label{trueStates3states}
  }
  \caption{{\bf Representation of hidden state creation for sequence $s$.} 
	If $D(\overline{S}_{t},S^*_{s,t})=0$ (in a), the true hidden state is ``no deviation'' (in c). If $D(\overline{S}_{t},S^*_{s,t})>0$ (in a), and no event occurs (in b), the true hidden state is ``context deviation'' (in c). If $D(\overline{S}_{t},S^*_{s,t})>0$ (in a) and an intraoperative adverse event occurs (in b), the true hidden state is ``event-deviation'' (in c).}
  \label{fig:TrueStatesRepresentation}
\end{figure}

In a first step, the HsMM is trained to define model $\lambda$ using observation sequences and true hidden sequences. These sequences, thus, need to be defined for each aligned iSPM. The observation sequences are defined as the concatenation of:
\begin{itemize}
	\item the three components of each aligned sequence's activities;
	\item the distance used to detect deviations between this sequence and the standard surgical process.
\end{itemize} 
We define three different hidden states: ``no deviation'' (compared to the standard surgical process), ``context deviation,'' and ``event deviation.'' These latter two are defined according to the definition given in Table~\ref{tab:deviations}.  In our case, we have not ``expert deviation'' due to the fact that all surgeries were performed by a single surgeon. Figure~\ref{fig:TrueStatesRepresentation} presents the creation of the true hidden sequences. If the distance between a specific surgery $s$ and the standard surgical process at instant \textit{t} is not null ($D(\overline{S}_{t},S^*_{s,t})>0$ in Figure~\ref{distToMean}) and an intraoperative adverse event (in Figure~\ref{event}) occurs, the true hidden state is defined as ``event deviation'' (in Figure~\ref{trueStates3states}). If the distance between a specific surgery and the standard surgical process at instant \textit{t} is not null ($D(\overline{S}_{t},S^*_{s,t})>0$), but there is no iAE, the true hidden state is defined as ``context deviation.'' Any other case corresponds to a ``no deviation'' true hidden state.

The training step of our HsMM was performed using the forward-backward algorithm developed by Yu and Kobayashi \cite{yu_efficient_2003} and the enabled creation of detection model $\lambda$.

\section{Results}
\label{Results}
Our method was validated using a leave-one-out cross-validation. The HsMM training was performed on all patients, except one. The remaining operation was sampled and aligned with the standard surgical process $\overline{S}$ in order to create an aligned test surgery $S^*_{test}$. We have computed the distance between $\overline{S}$ and $S^*_{test}$. With this distance and $S^*_{test}$, we have computed the true hidden state sequence and observation sequence, as explained in Section \ref{HsMM}. Deviations were detected by feeding the observation sequence to the trained detection model $\lambda$. Each model was evaluated by comparing the detected deviation sequence with the true hidden state sequence.

We investigated the impact of the sampling rate (Section \ref{subsec:creationRef}) on the results. To this end, we varied the sampling rates between 2 to 12 samples-per-second in 1-second steps. We also studied the results at 12.5 samples-per-second, given that this sampling rate corresponds to half of the video frequency (25Hz).


The distribution of observations between each hidden state is very heterogeneous  (Table~\ref{tab:heterogeneity}): 68\% of them belonging to the ``no deviation'' state, 26\% to the ``context deviation'' state, and only 6\% to the ``event deviation'' state. Moreover, this distribution is also very heterogeneous  between surgeries especially for ``event deviation'' state, with a standard deviation of 6.80\% and a large range (minimum of 0\% and a maximum superior of 18\%). Due to these heterogeneities, we could not be satisfied with accuracy only, as performance metrics. Given the small amount of ``event deviation'' occurrences, we can reach an accuracy of 94\% if all observations belonging to the ``no deviation'' and ``context deviation'' states are correctly classified, even if none of the ``event deviation'' states were detected. We, thus, used recall and precision to accurately estimate our model's ability to classify event deviation types. All results are given with a confidence interval (CI) of 95\%.

\begin{table*}[!ht]
	
	\centering
	\caption{{\bf Distribution of observations for each hidden states.} }

    \begin{tabular}{|c|c|c|c|c|c|}\hline
    	&Mean (\%)&STD(\%)&Median(\%)&Min(\%)&Max(\%)\\\hline
        No deviation&	68.41&	4.22&	69.52&	60.05&	74.91\\\hline
		Context deviation &	25.86&	5.73&	26.91&	12.19&	32.74\\\hline
		Event deviation&	5.73&	6.80&	2.83&	0.00&	18.13\\\hline    
    \end{tabular}    			
  \label{tab:heterogeneity}
\end{table*}

Kendall's Tau, a non-parametric test, was performed to examine a possible statistical correlation between the sampling rate and accuracy, precision and recall for each hidden state. We chose a level of 0.05 to consider the correlation statistically significant. With seven statistical tests (one for accuracy, three for recall, and three for precision), we were in the context of multiple comparisons. To counteract the problem of false-positive results in multiple comparisons, we employed the Bonferroni correction method \cite{bonferroni_teoria_1936,dunn_multiple_1961}. Therefore, the statistical significance level was set at 0.0071 (0.05/7).

We additionally investigated the model errors more closely in order to understand the reasons underlying failed classification of deviation types, with the aim to classify event deviations (Section \ref{error_identification}).

\subsection{Validation of the standard surgical process}\label{validationStdSurgPro}
The deviation detection method proposed is based on the learning phase of our HsMM. It is thus dependent on the hidden state sequences defined, due to the distance between a specific surgery and the standard surgical process (Figure \ref{distToMean}). It is essential to validate the consistency of the standard surgical process (section \ref{creationSSP}) in terms of surgical workflow, i.e. the sequence of computed activities that could happen in a real sequence. If the standard surgical process is not consistent, such as closure of the body occurring before the first skin incision, the deviation detection will not be correct.
To carry out this validation, we computed multiple standard surgical processes. One was computed with all iSPMs available, and the others by removing one or more iSPMs before computation. We have randomly shown to two surgeons graphical representations of real surgeries (iSPMs) and graphical representations of standard surgical processes. Several examples are available as supplementary material. The surgeons were asked if they deemed that, according to their surgical expertise, each representation was consistent in terms of surgical workflow, and whether each representation corresponded to either real surgery or computed standard surgical process. They considered all representations consistent, being unable to distinguish standard surgical processes from real surgeries.
With this validation, based on experts’ opinions, we assessed the computation of standard surgical processes produces realistic sequences in terms of surgical workflow.

\subsection{Deviation classification results}
\label{observations_Results}
\begin{figure}[h]

\centering
 \subfloat[Accuracy]{%
    \includegraphics[height=3.5cm]{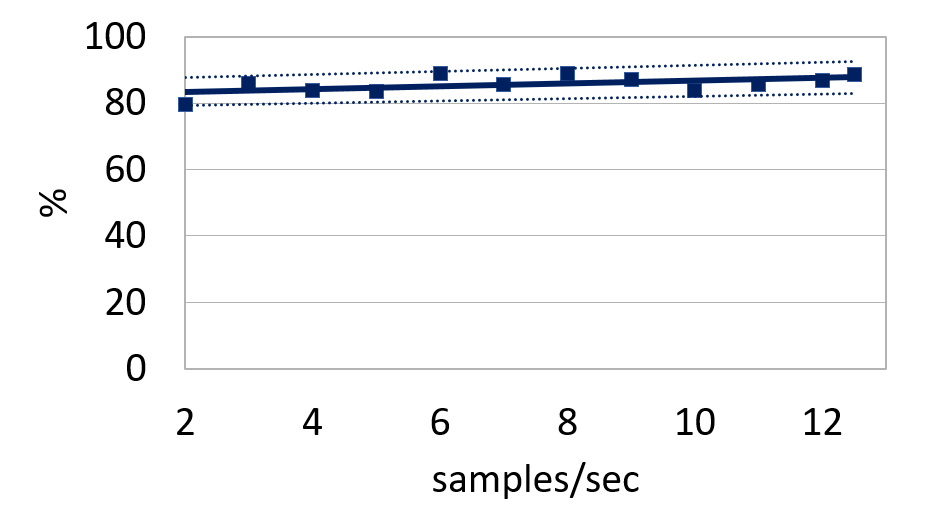}%
    \label{NDvsSDvsED-accuracy}
		
 }
 \subfloat[Recall]{
    \includegraphics[height=3.5cm]{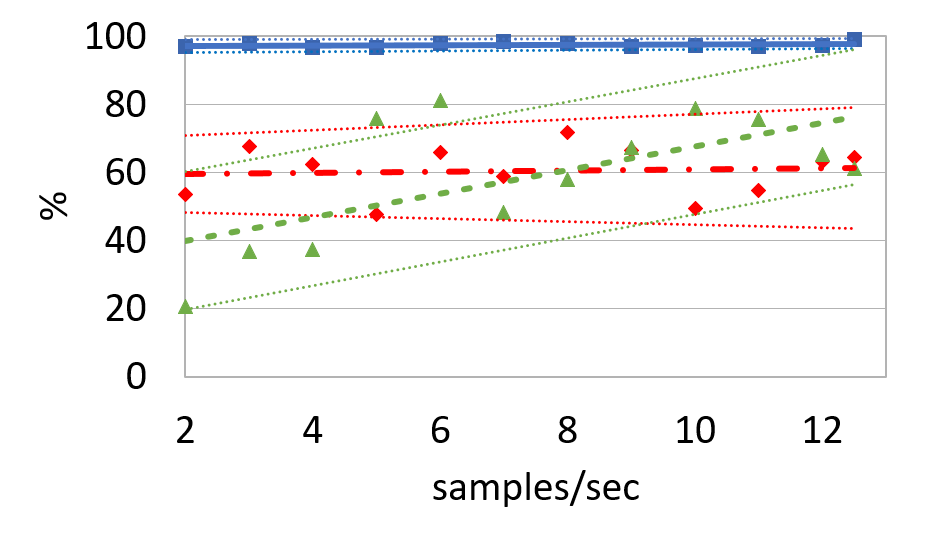}
    \label{NDvsSDvsED-PC}
 }

 \subfloat[Precision]{
    \includegraphics[height=3.5cm]{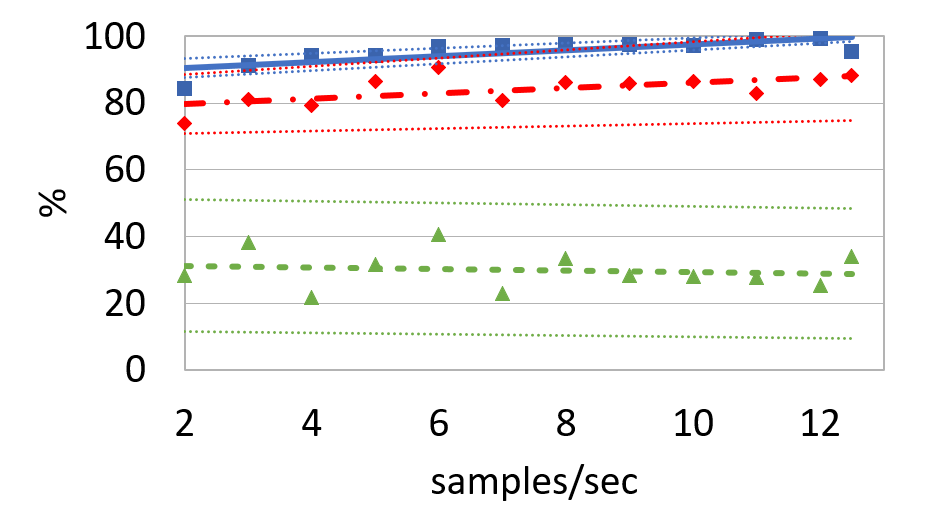}
    \label{NDvsSDvsED-PV}
 }
 \caption{{\bf Graph of the classification of surgical deviations.}
 Dotted lines represent the trend curve of the 95\% confidence intervals. In Figures~(b) and (c), the blue solid line corresponds to the ``no deviation'' state, the red dotted-solid line to the ``context deviation'' state, and the green dashed line to the ``event deviation'' state.
 }
 \label{fig:NDvsSDvsED}

\end{figure}

Figure~\ref{fig:NDvsSDvsED} and Table~\ref{tab:NDvsSDvsED} present the results of the deviation classification for the three metrics (accuracy, recall, and precision) at different sampling rates. A table providing all results is available as supplementary material.   

\begin{table*}[!ht]
	
	\centering
	\caption{{\bf Results of the classification of surgical deviations.} }

    \begin{tabular}{|c|c|c|c|c||c|c|}\hline
    	\multicolumn{2}{|c|}{Samples/sec}& 2 & 8 & 12 & tau & p-value\\\hline
    	\multicolumn{2}{|c|}{Accuracy(\%)}& 75.69 - 83.65 & 84.63 - 93.43 & 82.31 - 91.23 & 0.2727 &0.1248\\\hline
    	\multirow{3}{*}{\rotatebox{90}{Recall}}& ND(\%)&94.67 - 99.29 & 96.40 - 99.34& 95.67 - 99.21&0.24.24 &0.1554\\\cline{2-7}
    	&CD(\%)&42.39 - 64.77& 53.92 - 89.66& 46.87 - 79.33&0.0606 &0.4203\\\cline{2-7}
    	&ED(\%)&11.32 - 29.72& 33.09 - 82.53& 42.38 - 88.32&0.3636 &0.0580\\\hline
    	\multirow{3}{*}{\rotatebox{90}{Precision}}&ND(\%)&80.41 - 88.43&96.19 - 99.19& 98.83 - 99.56&0.6667 & 0.0009*\\\cline{2-7}
    	&CD(\%)&61.04 - 86.68 & 78.21 - 94.33& 71.97 - 100&0.4848 &0.0155\\\cline{2-7}
    	&ED(\%)&10.38 - 46.46& 9.73 - 57.29& 13.33 - 37.05&-0.0909&0.6808\\\hline
    
    \end{tabular}

    \begin{flushleft}  ND: ``no deviation.'' CD: ``context deviation.'' ED: ``event deviation.'' The star (*) represents a significant relationship between the sampling rate and results.
		\end{flushleft}
		
  \label{tab:NDvsSDvsED}
\end{table*}

\FloatBarrier

The accuracy ranged between 80 \% and 90 \% with a one-half the width of the CI inferior to 6 \% for two samples-per-second or more. We could observe a non-significant (p-value=0.1248) upward trend in accuracy along with the sampling rate (Figure~\ref{NDvsSDvsED-accuracy}).

\modif{The recall for the ``no deviation'' state ranged between 96 \% and 99 \% with a CI range inferior to 5 \% for all sampling rates}. On the other hand, the recall for the ``context deviation'' state fluctuated between a mean of 50 \% and 72 \% without any specific trend, however the CI range (red dotted line) trend to disperse with the sampling rate. For the ``event deviation'' state, the sampling rate appeared to impact the recall (green dashed line in Figure~\ref{NDvsSDvsED-PC}). For two samples-per-second, the recall as a 95\% CI of [11.32 - 29.72], whereas it was [42.38 - 88.32] for 12 samples-per-second. However, this trend proved to be not significant (p-value=0.0580) . Results are very different between sequences as we could shown with the CI.

The precision for the ``no deviation'' state significantly (p-value = 0.0009) increased from [80.41 - 88.43] to [98.83 - 99.21]  with the sampling rate and a half CI range inferior to 6 \%. The precision for the ``context deviation'' state fluctuated between a mean of 73 \% and 98 \% with a half CI range inferior to 15 \%, except for 12.5 samples-per-second where the half CI range is 21.16 \%, the trend being statistically not significant (p-value=0.0155). \modif{For the ``event deviation'' state, the mean precision was less than 40 \% for all sampling rates}. As for the recall, the CI demonstrate a high variability between sequences for the precision of ``event deviation'' state, with for example result between [9.73 - 57.29] for 8 samples per second.

\subsection{Analysis of the model's errors in classifying event deviations}
\label{error_identification}
To understand why the mean precision for ``event deviation'' state was less than to 40 \%, we analyzed the classification errors. We noticed that over 98 \% of the observations falsely classified as ``event deviation'' were actually representative of  ``context deviation.'' These observations were then classified into four categories:

\begin{enumerate}
 \item Rarely wrongly classified: less than 1\% of observations belonging to this observation type (same action verb, surgical instrument, anatomic target, and distance $D$) were wrongly classified by the model.
 \item Untrained: this observation type was not present in the training dataset.
 \item Correctly trained: in the training dataset, this observation type was characteristic of the ``event deviation'' state.
 \item Other: observations that did not belong to the previous three categories.
\end{enumerate}

\begin{figure}[h]
\centering
     \includegraphics[height=3.5cm]{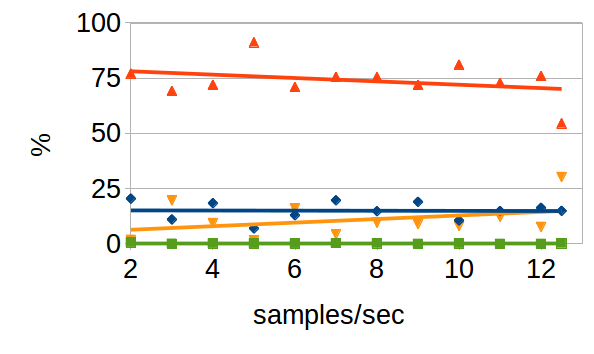}     
     \caption{{\bf Graphical distribution of the four categories of falsely classified observations for different sampling rates.} The green line corresponds to the ``rarely wrongly classified'' category, blue line to the ``untrained'' category, yellow line to the ``correctly trained'' category, and red line to the ``other'' category.}
     \label{fig:NDvsSDvsED-FalseED}
\end{figure}

The distribution of the observations within of these four categories has been provided in Figure~\ref{fig:NDvsSDvsED-FalseED}. Category 1 observations (rarely wrongly classified) were negligible for all sampling rates, accounting for less than 1 \% of errors. Category 2 (untrained) and 3 (correctly trained) observations accounted for less than 20 \% of falsely-classified observations, with a downward trend in the sampling rate for category 2, and upward trend for category 3 (Figure~\ref{fig:NDvsSDvsED-FalseED}). Category 4 (other) represented more than 50 \% of the falsely-classified observations, though we observed a downward trend with increasing sampling rates.

\section{Discussion}
Our validation study's results have clearly shown the high accuracy of our approach in detecting deviations, for all sampling rates, during the surgical process. However, the distinction between the deviation types does not prove to be effective, and results have high variability between sequences. Our model classifies a number of "context deviation" states as "event deviation" states. However, from a patient's safety point of view, it proves crucial that deviations due to intraoperative events are not missed, even if false detections do occur.

We have deeply analyzed the misclassification between the two deviation types (Figure~\ref{fig:NDvsSDvsED-FalseED}). Based on this analysis, four observation categories were extracted, with each category interpreted as follows:
\begin{itemize}
    \item Category 1 errors (rarely wrongly detected observations) are caused by the time taken by  the model  to perform a state transition;
    \item Category 2 errors (untrained observations) are due to a lack of data: because the model did not encounter the observation in the training phase, it provides an arbitrary result. A larger dataset might reduce the number of observations pertaining to this category; 
	\item Category 3 errors (correctly trained observations) could be caused by activities that would have led to an ``event deviation,'' as observed in the training dataset, but were corrected by the surgeon before they occurred. A study of surgical behavior is warranted to confirm this hypothesis;
	\item a large number of observations were classified as Category 4 (other). Our interpretation for this error type is that surgical activities, such as defined today (action verb, surgical instrument, and anatomic target), may not capture enough information concerning the surgical scene. To solve this issue, a more refined description of activities appears necessary. For instance, by combining the laparoscopic images with registered complementary per-operative imaging modalities, such as ultrasounds \cite{lanchon_augmented_2016,billings_system_2012} or fluorescence imaging \cite{voros_devices_2013}, it may be easier to identify vasculature information and express the distance between the surgical instrument and this underlying vasculature as a label (too close, close or safe). This information could be treated similarly to any activity component in our approach. 
\end{itemize}

The CI range demonstrates a high variability of the results between sequences, especially for ``event deviation'' results. This could be explained by the high variability on the distribution of each type of deviation, as shown in Table~\ref{tab:heterogeneity}, and the limited size of our database. Indeed, when we test the sequence with the more important distribution of event deviations, we have a higher probability to encounter observations never encounter during the training phase.

Our study presents several limitations. First, the entire study has been based on manual annotations performed by one observer only. Neumuth et al. \cite{neumuth_validation_2009} studied the reliability of the annotation process and concluded that ``granularity was reconstructed correctly by 90\%, content by 91\%, and the mean temporal accuracy was 1.8 s." This temporal variability is reinforced by Huaulm\'e et al. \cite{huaulme_automatic_2019} which demonstrated a temporal inter-variability which could have a relative standard deviation superior to 18\% for activities when multiple observers are implied. According to the literature, we could thus consider having similar results, i.e. if we implied several observers, the identified activities will be very similar but we will introduce temporal variability. As explained in section \ref{MD_NLTS}, this works was based on activity sequences rather than their durations. Consequently, this variability would not be relevant information for our method.

Second, although we have validated the consistency of the standard surgical process $\overline{S}$ (Section \ref{creationSSP}), in terms of surgical workflow, its creation can be a source of false deviation classification. Remember, we have determined the activity at each instant \textit{t} by selecting the most frequent activity in the aligned sequences. However, if the aligned sequences are too heterogeneous at one instant, this most frequent activity might not be present in the majority of aligned sequences (e.g. it might only represent a small percentage of the activities at time t, even though it is the most frequent activity). For further developments of this approach, it will be paramount to consider the probability of the chosen activity or allow for alternative surgical paths within the standard surgical process.

The third limitation concerns the choice of only considering the dominant hand of the surgeon for classifying deviations. It could be of interest to add the information provided by the non-dominant hand to further investigate its influence on deviation detection. To this end, the activities of the second hand must be annotated, while our method has to be improved by modifying our Multi-Dimensional Non-Linear Temporal Scaling method and using a coupled Hidden semi-Markov Model with two observation sequences (one per hand), and one hidden state sequence.
	
Finally, our dataset comprises of surgeries performed by a single surgeon, while we have not considered differing habits or expertise levels among the surgeons. We have, therefore, removed one level of complexity. Furthermore, the dataset includes bleeding events only. We cannot predict the performance of our method for other types of iAEs. However, our annotation methodology, which only relies on activity annotations, would be identical whatever the type of iAE, as long as the iAE start and end times can be identified by the expert surgeons involved in the annotation process. Moreover, we did not take into account at this stage organizational/context factors into our annotation process. Indeed, operating room disruptions due to e.g. the composition of the surgical team could also result in deviations. They are currently considered as ``operating room disruptions" and belong to context deviations. Future works would examine the robustness of our approach by including more types of iAEs, multi-surgeon data to include the expert deviation classification and study the influence of the surgical team composition on deviation occurrences. Collect data from multiple surgeons and other surgical team members will allow using more complex approaches than a simple ``leave-one-out" one, e.g. a ``leave-one-user-out" or one where the couple surgeon/assistant is excluded from the training.

\section{Conclusion}
Surgical deviation classification is challenging and should enable us to understand the hidden processes underlying their occurrence. We have, herein, proposed the first offline method for automatically classifying deviations based on their type (event deviation or context deviation). The method, namely Multi-Dimensional Non-Linear Temporal Scaling followed by a Hidden semi-Markov Model, has provided interesting initial results, whereas its precision still needs to be improved.

The detection of event deviations is an important preliminary step towards the identification of iAEs. Indeed, event deviations are a marker of the occurrence of one or multiples iAEs. This could help determine the exact moment when iAEs occur. Moreover, the objective of an event deviation is to ``correct or limit the impacts of iAEs'' (Table~\ref{tab:deviations}), so by studying the anatomical structure concerned by event deviations, it will be possible to determine which one is impacted. Of course, to make a complete identification of iAEs, further work will be necessary.

To propose routine surgery applications of our method, two further improvements are required. The first is to develop an on-line multi-dimensional alignment method. Recently, Forestier \textit{et al.} \cite{forestier_optimal_2015} proposed a method designed to create an online one-dimensional alignment. The second aspect pertains to creating a reliable and automatic online activity recognition method \cite{katic_ontology-based_2013,despinoy_unsupervised_2015,dergachyova_automatic_2016}. With these two developments available, a real-time implementation of our method will be rendered possible.

\section*{Conflict of interest statement}
The authors declare that they have no conflict of interest.

\section*{Acknowledgements}
\noindent
This work was partially supported by French state funds managed by the ANR within the Investissements d'Avenir programme (Labex CAMI) under reference ANR-11-LABX-0004.

\noindent
Authors thanks the IRT b\textless \textgreater com for the provision of the software ``Surgery Workflow Toolbox [annotate]'' , used for this work.

\bibliographystyle{elsarticle-num}
\bibliography{Detection_EI}   

\begin{thebibliography}{10}
\expandafter\ifx\csname url\endcsname\relax
  \def\url#1{\texttt{#1}}\fi
\expandafter\ifx\csname urlprefix\endcsname\relax\def\urlprefix{URL }\fi
\expandafter\ifx\csname href\endcsname\relax
  \def\href#1#2{#2} \def\path#1{#1}\fi

\bibitem{anderson_surgical_2013}
O.~Anderson, R.~Davis, G.~B. Hanna, C.~A. Vincent, Surgical adverse events: a
  systematic review, The American Journal of Surgery 206~(2) (2013) 253--262.
\newblock \href {http://dx.doi.org/10.1016/j.amjsurg.2012.11.009}
  {\path{doi:10.1016/j.amjsurg.2012.11.009}}.

\bibitem{world_health_organization_who_2005}
{World Health Organization}, {WHO} draft guidelines for adverse event reporting
  and learning systems (2005).

\bibitem{jannin_modeling_2001}
P.~Jannin, M.~Raimbault, X.~Morandi, B.~Gibaud, Modeling {Surgical}
  {Procedures} for {Multimodal} {Image}-{Guided} {Neurosurgery}, in: W.~J.
  Niessen, M.~A. Viergever (Eds.), Medical {Image} {Computing} and
  {Computer}-{Assisted} {Intervention} – {MICCAI} 2001, no. 2208 in Lecture
  {Notes} in {Computer} {Science}, Springer Berlin Heidelberg, 2001, pp.
  565--572.

\bibitem{lalys_surgical_2013}
F.~Lalys, P.~Jannin, Surgical process modelling: a review, International
  Journal of Computer Assisted Radiology and Surgery 9~(3) (2013) 495--511.

\bibitem{riffaud_recording_2010}
L.~Riffaud, T.~Neumuth, X.~Morandi, C.~Trantakis, J.~Meixensberger, O.~Burgert,
  B.~Trelhu, P.~Jannin, Recording of {Surgical} {Processes}: {A} {Study}
  {Comparing} {Senior} and {Junior} {Neurosurgeons} {During} {Lumbar} {Disc}
  {Herniation} {Surgery}., Operative Neurosurgery 67 (2010) ons325--ons332.

\bibitem{forestier_classification_2012}
G.~Forestier, F.~Lalys, L.~Riffaud, B.~Trelhu, P.~Jannin, Classification of
  surgical processes using dynamic time warping, Journal of Biomedical
  Informatics 45~(2) (2012) 255--264.
\newblock \href {http://dx.doi.org/10.1016/j.jbi.2011.11.002}
  {\path{doi:10.1016/j.jbi.2011.11.002}}.

\bibitem{sandberg_deliberate_2005}
W.~S. Sandberg, B.~Daily, M.~Egan, J.~E. Stahl, J.~M. Goldman, R.~A. Wiklund,
  D.~Rattner, Deliberate {Perioperative} {Systems} {Design} {Improves}
  {Operating} {Room} {Throughput}:, Anesthesiology 103~(2) (2005) 406--418.
\newblock \href {http://dx.doi.org/10.1097/00000542-200508000-00025}
  {\path{doi:10.1097/00000542-200508000-00025}}.

\bibitem{padoy_-line_2008}
N.~Padoy, B.~Tobias, H.~Feussner, M.-O. Berger, N.~Navab, On-line {Recognition}
  of {Surgical} {Activity} for {Monitoring} in the {Operating} {Room}, 2008,
  pp. 1718--1724.

\bibitem{ko_surgery_2007}
S.-Y. Ko, J.~Kim, W.-J. Lee, D.-S. Kwon, Surgery task model for intelligent
  interaction between surgeon and laparoscopic assistant robot, International
  Journal of Assitive Robotics and Mechatronics 8~(1) (2007) 38--46.

\bibitem{nomm_recognition_2008}
S.~Nomm, E.~Petlenkov, J.~Vain, J.~Belikov, F.~Miyawaki, K.~Yoshimitsu,
  Recognition of the surgeon’s motions during endoscopic operation by
  statistics based algorithm and neural networks based {ANARX} models, Proc Int
  Fed Automatic Control 17~(1).

\bibitem{bouarfa_workflow_2012}
L.~Bouarfa, J.~Dankelman, Workflow mining and outlier detection from clinical
  activity logs, Journal of Biomedical Informatics 45~(6) (2012) 1185--1190.
\newblock \href {http://dx.doi.org/10.1016/j.jbi.2012.08.003}
  {\path{doi:10.1016/j.jbi.2012.08.003}}.

\bibitem{jadhav_anomaly_2013}
S.~N. Jadhav, K.~Bhandari, Anomaly {Detection} {Using} {Hidden} {Markov}
  {Model}, International Journal of Computational Engineering Research (IJCER)
  (2013) 28.

\bibitem{tan_hidden_2008}
X.~Tan, H.~Xi, Hidden semi-{Markov} model for anomaly detection, Applied
  Mathematics and Computation 205~(2) (2008) 562--567.
\newblock \href {http://dx.doi.org/10.1016/j.amc.2008.05.028}
  {\path{doi:10.1016/j.amc.2008.05.028}}.

\bibitem{forestier_non-linear_2014}
G.~Forestier, F.~Petitjean, L.~Riffaud, P.~Jannin, Non-linear temporal scaling
  of surgical processes, Artificial Intelligence in Medicine 62~(3) (2014)
  143--152.
\newblock \href {http://dx.doi.org/10.1016/j.artmed.2014.10.007}
  {\path{doi:10.1016/j.artmed.2014.10.007}}.

\bibitem{clark_cognitive_2008}
R.~E. Clark, D.~F. Feldon, J.~J.~G. van Merriënboer, K.~A. Yates, S.~Early,
  Cognitive task analysis, Handbook of research on educational communications
  and technology 3 (2008) 577--593.

\bibitem{garraud_ontology-based_2014}
C.~Garraud, B.~Gibaud, C.~Penet, G.~Gazuguel, G.~Dardenne, P.~Jannin, An
  {Ontology}-based {Software} {Suite} for the {Analysis} of {Surgical}
  {Process} {Model}., in: Proceedings of {Surgetica}'2014, Chambery, France,
  2014, pp. 243--245.

\bibitem{petitjean_global_2011}
F.~Petitjean, A.~Ketterlin, P.~Gançarski, A global averaging method for
  dynamic time warping, with applications to clustering, Pattern Recognition
  44~(3) (2011) 678--693.
\newblock \href {http://dx.doi.org/10.1016/j.patcog.2010.09.013}
  {\path{doi:10.1016/j.patcog.2010.09.013}}.

\bibitem{shokoohi-yekta_non-trivial_2015}
M.~Shokoohi-Yekta, J.~Wang, E.~Keogh, On the {Non}-{Trivial} {Generalization}
  of {Dynamic} {Time} {Warping} to the {Multi}-{Dimensional} {Case}, in: Data
  {Mining}. {Proceeding} of the 2015 {International} {Conference} on, SIAM,
  2015, pp. 39--48.

\bibitem{yu_efficient_2003}
S.-Z. Yu, H.~Kobayashi, An efficient forward-backward algorithm for an
  explicit-duration hidden {Markov} model, IEEE Signal Processing Letters
  10~(1) (2003) 11--14.
\newblock \href {http://dx.doi.org/10.1109/LSP.2002.806705}
  {\path{doi:10.1109/LSP.2002.806705}}.

\bibitem{bonferroni_teoria_1936}
C.~E. Bonferroni, Teoria statistica delle classi e calcolo delle probabilita,
  Libreria internazionale Seeber, 1936.

\bibitem{dunn_multiple_1961}
O.~J. Dunn, Multiple {Comparisons} {Among} {Means}, Journal of the American
  Statistical Association 56~(293) (1961) 52.
\newblock \href {http://dx.doi.org/10.2307/2282330}
  {\path{doi:10.2307/2282330}}.

\bibitem{lanchon_augmented_2016}
C.~Lanchon, G.~Custillon, A.~Moreau-Gaudry, J.-L. Descotes, J.-A. Long,
  G.~Fiard, S.~Voros, Augmented {Reality} {Using} {Transurethral} {Ultrasound}
  for {Laparoscopic} {Radical} {Prostatectomy}: {Preclinical} {Evaluation}, The
  Journal of urology 196~(1) (2016) 244--250.

\bibitem{billings_system_2012}
S.~Billings, N.~Deshmukh, H.~Kang, R.~Taylor, E.~M. Boctor, System for
  robot-assisted real-time laparoscopic ultrasound elastography, in: {SPIE}
  {Medical} {Imaging}, International Society for Optics and Photonics, 2012,
  pp. 83161W--83161W.

\bibitem{voros_devices_2013}
S.~Voros, A.~Moreau-Gaudry, B.~Tamadazte, G.~Custillon, R.~Heus, M.-P.
  Montmasson, F.~Giroud, O.~Gaiffe, C.~Pieralli, G.~Fiard, J.-A. Long, J.-L.
  Descotes, C.~Vidal, A.~Nguyen-Dinh, P.~Cinquin, Devices and systems targeted
  towards augmented robotic radical prostatectomy, IRBM 34~(2) (2013) 139--146.
\newblock \href {http://dx.doi.org/10.1016/j.irbm.2013.01.014}
  {\path{doi:10.1016/j.irbm.2013.01.014}}.

\bibitem{neumuth_validation_2009}
T.~Neumuth, P.~Jannin, G.~Strauss, J.~Meixensberger, O.~Burgert, Validation of
  {Knowledge} {Acquisition} for {Surgical} {Process} {Models}, Journal of the
  American Medical Informatics Association 16~(1) (2009) 72--80.
\newblock \href {http://dx.doi.org/10.1197/jamia.M2748}
  {\path{doi:10.1197/jamia.M2748}}.

\bibitem{huaulme_automatic_2019}
A.~Huaulm\'e, F.~Despinoy, S.~A.~H. Perez, K.~Harada, M.~Mitsuishi, P.~Jannin,
  Automatic annotation of surgical activities using virtual reality
  environments, International Journal of Computer Assisted Radiology and
  Surgery\href {http://dx.doi.org/10.1007/s11548-019-02008-x}
  {\path{doi:10.1007/s11548-019-02008-x}}.

\bibitem{forestier_optimal_2015}
G.~Forestier, F.~Petitjean, L.~Riffaud, P.~Jannin, Optimal {Sub}-{Sequence}
  {Matching} for the {Automatic} {Prediction} of {Surgical} {Tasks}, in:
  Artificial {Intelligence} in {Medicine}, Vol. 9105, Springer International
  Publishing, Cham, 2015, pp. 123--132.

\bibitem{katic_ontology-based_2013}
D.~Katić, A.-L. Wekerle, F.~Gärtner, H.~Kenngott, B.~P. Müller-Stich,
  R.~Dillmann, S.~Speidel, Ontology-based prediction of surgical events in
  laparoscopic surgery, 2013, p. 86711A.
\newblock \href {http://dx.doi.org/10.1117/12.2007895}
  {\path{doi:10.1117/12.2007895}}.

\bibitem{despinoy_unsupervised_2015}
F.~Despinoy, D.~Bouget, G.~Forestier, C.~Penet, N.~Zemiti, P.~Poignet,
  P.~Jannin, Unsupervised trajectory segmentation for surgical gesture
  recognition in robotic training, IEEE Transactions on Biomedical Engineering
  63~(6) (2015) 1280--1291.

\bibitem{dergachyova_automatic_2016}
O.~Dergachyova, D.~Bouget, A.~Huaulm\'e, X.~Morandi, P.~Jannin, Automatic
  data-driven real-time segmentation and recognition of surgical workflow,
  International Journal of Computer Assisted Radiology and Surgery\href
  {http://dx.doi.org/10.1007/s11548-016-1371-x}
  {\path{doi:10.1007/s11548-016-1371-x}}.

\end{thebibliography}

\end{document}